\documentclass{llncs}
\pdfoutput=1 
\usepackage[utf8]{inputenc}
\usepackage{graphicx}
\usepackage{amsmath}
\usepackage{tikz-uml}
\usepackage{tikz}
\usepackage{relsize}
\usepackage[export]{adjustbox}
\usepackage{hyperref}
\usepackage{subcaption}
\usepackage{url}
\usepackage{ulem}
\usepackage{subfiles}
\usepackage{standalone}
\usepackage{xspace}
\usepackage{booktabs}
\usepackage{amssymb,amsmath}
\usetikzlibrary{decorations.pathmorphing}
\usetikzlibrary{%
  arrows,%
  shapes.misc,
  shapes.arrows,%
  chains,%
  matrix,%
  positioning,
  scopes,%
  decorations.pathmorphing,
  shadows%
}

\newcommand\CC{C\nolinebreak[4]\hspace{-.05em}\raisebox{.4ex}{\relsize{-2}{\textbf{++}}}\xspace}
\newcommand\nova{NOvA\xspace}

\title{CUDA Support in GNA Data Analysis Framework}
\author{Anna Fatkina \and  Maxim Gonchar \and Liudmila Kolupaeva  \and Dmitry Naumov \and  Konstantin Treskov }
\institute {Joint Institute for Nuclear Research,\newline Joliot-Curie, 6, Dubna, Moscow region, Russia, 141980 }
\date{April 2018}

\begin{document}

\maketitle

\begin{abstract}
Usage of GPUs as co-processors is a well-established approach to accelerate costly algorithms operating on matrices and
vectors.

We aim to further improve the performance of the Global Neutrino Analysis framework (GNA) by adding GPU support in a way that is transparent to the end user. To achieve our goal we use CUDA, a state of the art technology providing GPGPU programming
methods.

In this paper we describe new features of GNA related to CUDA support. Some specific framework features that influence
GPGPU integration are also explained. The paper investigates the feasibility of GPU technology application
and shows an example of the achieved acceleration of an algorithm implemented within framework. Benchmarks show a significant performance
increase when using GPU transformations.

The project is currently in the developmental phase. Our plans include implementation of the set of transformations
necessary for the data analysis in the GNA framework and tests of the GPU expediency in the complete analysis chain.
\end{abstract}

\keywords{CUDA, GPGPU, parallel computing, data analysis, neutrino}

\section{Introduction}

The neutrino is weakly interacting neutral fermion.
There are three types of these particles $\nu_1$, $\nu_2$ and $\nu_3$ with masses $m_1$, $m_2$ and
$m_3$, respectively.
These particles interact with charged leptons (electron, muon and tau) with interaction strengths
determined by elements $V_{\alpha i}$ of the lepton mixing matrix $V$, named after
Pontecorvo-Maki-Nakagawa-Sakata.

Two facts, that neutrino masses are all different and that $V$ is not a diagonal matrix, lead to a
spectacular quantum mechanical phenomenon known as neutrino oscillations.
Its firm experimental confirmation was celebrated by the 2015 Nobel Prize in physics and the 2016
Breakthrough Prize in Fundamental Physics~\cite{2016NuPhB1O,kajita2016nobel,mcdonald2016nobel}.

Neutrino physics entered the stage of precision measurements and addressing remained open questions:
neutrino mass hierarchy, if neutrino is  Majorana particle, and others.
Both require an accurate, fast and flexible tool for a combined analysis of neutrino world data.
Our team began a development of the corresponding software GNA based on our experience in Daya
Bay~\cite{An:2016ses} (`Analysis D'), JUNO~\cite{An:2015jdp} and \nova~\cite{Adamson:2017gxd}
experiments.

GNA is an universal tool for building comprehensive physical models and statistical data analysis,
designed with neutrino experiments in mind. It was initially created as software for the JUNO and Daya
Bay experiments in a flexible and
efficient way.
The name GNA stands for Global Neutrino Analysis, as the package
introduces tools for the combined analysis of the physical data. The framework is described in more
detail in the following section.

GPUs (Graphics Processing Units) are used today for a much wider range of problems than simply
processing graphics, including data analysis in science~\cite{al2010gpu,fatkina2017application}.
Video cards can be used as co-processors on both personal computers and
high-performance servers. There exist free tools that provide an interface for GPU programming such
as CUDA~\cite{nickolls2008scalable}, OpenACC~\cite{farber2016parallel} or
OpenCL~\cite{howes2015opencl}.

We have added CUDA support to the GNA framework in order to achieve better performance during the processing of
vector data. With this architecture the input data is mapped on multiple threads that are executed in
parallel. Because a GPU platform has hundreds of times more threads compared to modern CPUs it is
especially suitable for running data-parallel algorithms.

The CUDA Toolkit is developed by NVIDIA and supports only NVIDIA graphics accelerators. This narrows the
range of compatible acceleration devices compared to other tools. Nevertheless, the CUDA Toolkit
provides a number optimized numerical routines. Also, NVIDIA GPUs are
quite popular and are widely used in common desktop computers and laptops.

It this paper we describe the way in which CUDA is integrated in GNA, and its implications from both the
end-user and developer points of view. Major implementation details
are discussed. A review of our future plans for GPU-based development is also presented.

\section{GNA Architecture}

The computation process in GNA is represented by a directed graph in which nodes represent functions and edges
present the data flow. Nodes are called transformations, which is an abstraction layer for \CC functions. They may have
inputs (arguments) and have at least one output (return values). Transformations typically operate on data arrays.
A computational graph describes how transformations interact with each other. Because transformations are encapsulated and
have universal interfaces a high flexibility is achieved.

Data analysis in GNA consists of two stages:
\begin{enumerate}
    \item Configuration stage on which the computational graph is created.
    \item Computational stage on which graph is evaluated.
\end{enumerate}

In the first stage the transformation instances are created, and outputs and inputs are bound together. This step is done only
once within Python and is flexible, but may be inefficient. The actual calculation happens on the second step.
Calculations are done within compiled \CC code and are usually executed repeatedly.

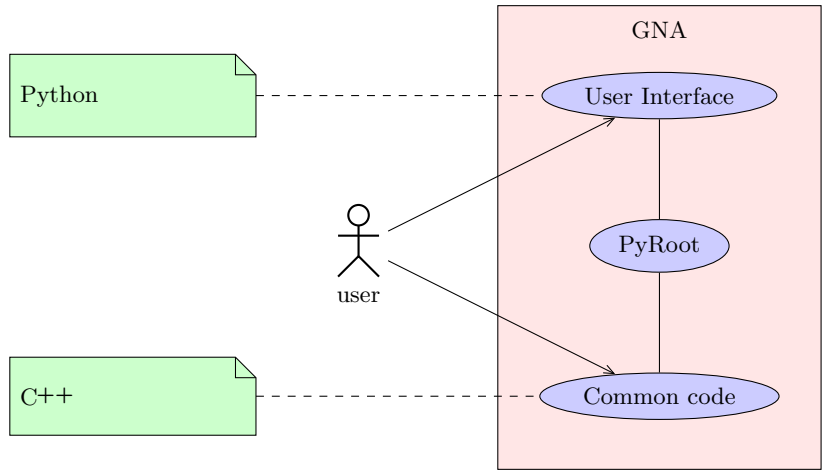
\begin{figure}[tb]
    \centering
    \begin{tikzpicture}
        \begin{umlsystem}[x=4, fill=red!10]{GNA}
        \umlusecase{User Interface}
        \umlusecase[y=-2]{PyRoot}
        \umlusecase[y=-4]{Common code}
        \end{umlsystem}

        \umlnote[x=-3]{usecase-1}{Python}
        \umlnote[x=-3, y=-4]{usecase-3}{\CC{}}

        \umlassoc{usecase-1}{usecase-2}
        \umlassoc{usecase-2}{usecase-3}

        \umlactor[y=-2]{user}

        \umluniassoc{user}{usecase-1}
        \umluniassoc{user}{usecase-3}
    \end{tikzpicture}
    \caption{GNA architecture schematic diagram.}
    \label{fig:mytikz}
\end{figure}

The generalized scheme of the framework is shown on Figure~\ref{fig:mytikz}. GNA has a Python user interface (UI) that is used
for building computation chains. The implementation of all transformations and the way they interact are described in
\CC{}. These two parts are linked via PyRoot.

The user may manage the computational process by using transformations already implemented in GNA. Transformations may also be
written by users themselves and added into the framework environment.

\subsection{Transformation}
A transformation is an encapsulated wrapper for a function that converts input data into output.

\begin{figure}
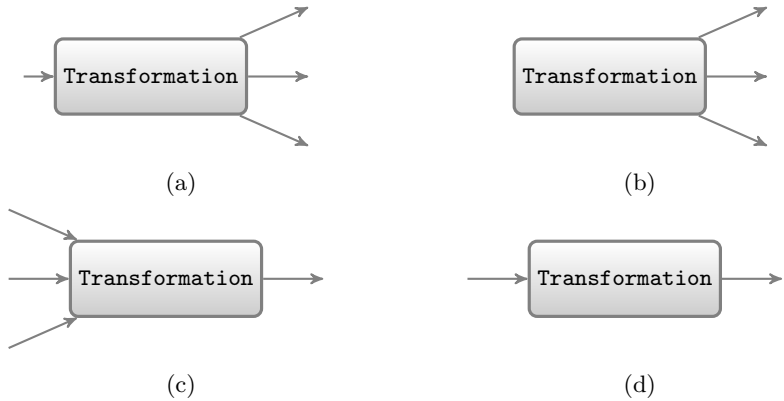

\centering
\begin{subfigure}{0.5\linewidth}
  \centering
  \input{trans-type1.tex}
  \caption{}
  \label{fig:test1}
\end{subfigure}%
\begin{subfigure}{0.5\linewidth}
  \centering
  \input{trans-type2.tex}
  \caption{}
  \label{fig:test2}
\end{subfigure}
\begin{subfigure}{0.5\linewidth}
  \centering
  \input{trans-type3.tex}
  \caption{}
  \label{fig:test3}
\end{subfigure}%
\begin{subfigure}{0.5\linewidth}
  \centering
  \input{trans-type4.tex}
  \caption{}
  \label{fig:test4}
\end{subfigure}
\caption{Example of transformation kinds.
    Intermediate transformation (\subref{fig:test1}) with a single input and multiple outputs.
    Initial transformation (\subref{fig:test2}) with multiple outputs.
    Intermediate transformation (\subref{fig:test3}) with multiple inputs and single output.
    Intermediate transformation (\subref{fig:test4}) with single input and single output.
  }
\label{fig:trasfs}
\end{figure}

Figure~\ref{fig:trasfs} schematically displays several kinds of transformations. Transformations may or
may not have inputs (marked by arrows on the left side) and must have at least one output (marked by arrows on the right
side). Inputs and outputs generally refer to data arrays. In addition to inputs transformation may also depend on
variables. A variable is a small input data type which usually refers to a single number.

Actual data is allocated on the transformation outputs.
Input data cannot be changed inside the transformation, it is a read-only state for the output it is
connected to. It enables us to ensure that data will not be modified by following transformations after it is computed.
A transformation is computed only once and the
result may be used multiple times afterwards. It will be re-computed only if any of the variables or inputs it depends on were modified. 

There is a set of predefined transformations implemented in the GNA framework. Because transformations are independent from each
other the set may be straightforwardly extended by the users. The guidelines on how to do this are provided in the framework
documentation~\cite{gnadoc}.

The typical computational chain that produces prediction for the reactor antineutrino experiments contains hundreds of
nodes and is evaluated within a time frame on the order of 0.1 seconds to seconds. The prediction is a histogram with 300 bins and
depends overall on 250 independent parameters. The prediction is then used in the process of multidimensional
minimization, which takes around 30 minutes for 15 free parameters or around 6 hours for all the model parameters, most
of which are constrained. Statistical analysis requires repeated minimization and may take several days to evaluate
confidence intervals. MC based methods, such as Feldman-Cousins, require millions of minimization procedures and may take
months when executed on a cluster. The framework is also suitable for building more complex graphs with evaluation times
on the order of seconds to hours.

\subsection{Computational graph}

\begin{figure}[h]
    \centering
    \tikzset{
  nonterminal/.style={
    rectangle, rounded corners=3pt,
    minimum size=10mm,
    very thick,
    draw=red!50!black!50,         
    top color=white,              
    bottom color=red!50!black!20, 
  },
  terminal/.style={
    rectangle, rounded corners=3pt,
    minimum size=10mm,
    very thick,draw=black!50,
    top color=white,bottom color=black!20,
    font=\ttfamily},
  crossred/.style={
    minimum size=10mm,
    very thick,draw=black!50, cross out,
    top color=white,bottom color=black!20,
    font=\ttfamily},
    cross/.style={
        append after command={
            [every edge/.append style={
                thick,
                red!50!black!50,
                shorten >=\pgflinewidth,
                shorten <=\pgflinewidth,
            }]
           (\tikzlastnode.north west) edge (\tikzlastnode.south east)
           (\tikzlastnode.north east) edge (\tikzlastnode.south west)
        }
    }
}

{
  \tikzset{terminal/.append style={text height=1.5ex,text depth=.25ex}}
  \tikzset{nonterminal/.append style={text height=1.5ex,text depth=.25ex}}
  \tikzset{cross/.append style={text height=1.5ex,text depth=.25ex}}
}

\begin{tikzpicture}[
        point/.style={coordinate},>=stealth',thick,draw=black!50,
        tip/.style={->,shorten >=0.007pt},every join/.style={rounded corners},
        hv path/.style={to path={-| (\tikztotarget)}},
        vh path/.style={to path={|- (\tikztotarget)}},
        text height=1.5ex,text depth=.25ex  
    ]
    \matrix[column sep=8mm, row sep=4mm] {
	\node (t1) [terminal] {$\mathrm{T_1}$}; & & \\
	\node (t2) [terminal] {$\mathrm{T_2}$}; & & \node (t4) [terminal] {$\mathrm{T_4}$}; & & \node (t5) [terminal] {$\mathrm{T_5}$}; & \\
	\node (t3) [terminal] {$\mathrm{T_3}$}; & & & & & \node (t8) [terminal] {$\mathrm{T_8}$};& & \node (p1) [point] {}; & \\
	&  & \node (t6) [terminal] {$\mathrm{T_6}$}; & & \node (t7) [terminal] {$\mathrm{T_7}$}; & & &\\
};
    { [start chain]
	\chainin (t8);
	{ [start branch=up]
        	\chainin (t4);
		{[start branch=out1]
          \chainin (t1) ;
          \chainin (t4) [join=by {tip,out=0,in=135}];
		}
        	{ [start branch=out2]
                	\chainin (t2) ;
			\chainin (t4) [join=by {tip}];
	        }
        	{ [start branch=out13]
                	\chainin (t3) ;
			\chainin (t4) [join=by {tip,out=0,in=-135}];
        	}
		\chainin (t5) [join=by {tip}];
		\chainin (t8) [join=by {tip,out=0,in=120}];
	}
	{ [start branch=down]
		\chainin (t6);
		{[start branch=down2]
			\chainin (t6);
			\chainin (t5) [join=by {tip,out=0,in=-110}];
		}
		{[start branch=down3	]
                        \chainin (t6);
                        \chainin (t8) [join=by {tip,out=0,in=180}];
                }
	        \chainin (t7) [join=by {tip}];
		\chainin (t8) [join=by {tip,out=0,in=-120}];
	}
	\chainin (p1) [join=by {tip}];

  }
\end{tikzpicture}
    \caption{Schematic example of the GNA computational graph.}
    \label{fig:comp-gr}
\end{figure}
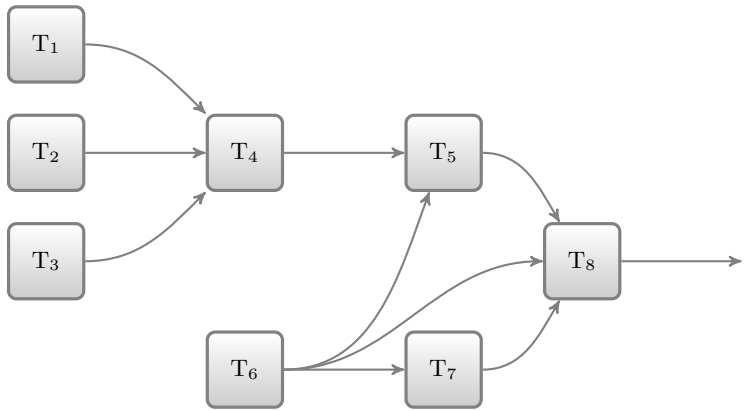

A computational graph is formed by a chain of transformations with inputs connected to outputs. Figure~\ref{fig:comp-gr} displays
a simple example of such a graph. This scheme shows that the same output may refer to and be referred by any number of
inputs. The graph may be
configured in an arbitrary way, as long as data types of the outputs are compatible with the requirements of the transformation they are connected to.

The graph is constructed using Python. Users describe the way transformations are chained via Python script or from the
command line interface. The result of any transformation may be read at any moment through the Python interface.

Lazy evaluation means that  the output of a transformation is computed on demand if the output is read by a caller.
In the case when the output of an intermediate transformation is accessed only preceding transformations are evaluated,
not the entire graph.

\subsection{Parallelism opportunities}

Parallel computing is a well-known method to speed up the computational process. There are methods to achieve performance
increases on different levels. The most efficient and safe method is to divide input data into smaller independent datasets
and execute the analysis on a distributed system~\cite{ballintijn2003proof,gankevich2017subord}. However, in real-world cases
analysis of those datasets often takes a long time. Due to this fact acceleration at an individual dataset level is also needed,
and may be implemented for multi-core CPUs or GPUs~\cite{iakushkin2017application}. In this paper we consider the
prospects for acceleration of computations in GNA on a framework level using GPGPU.

\begin{figure}[tb]
    \centering
    \includegraphics[width=\textwidth]{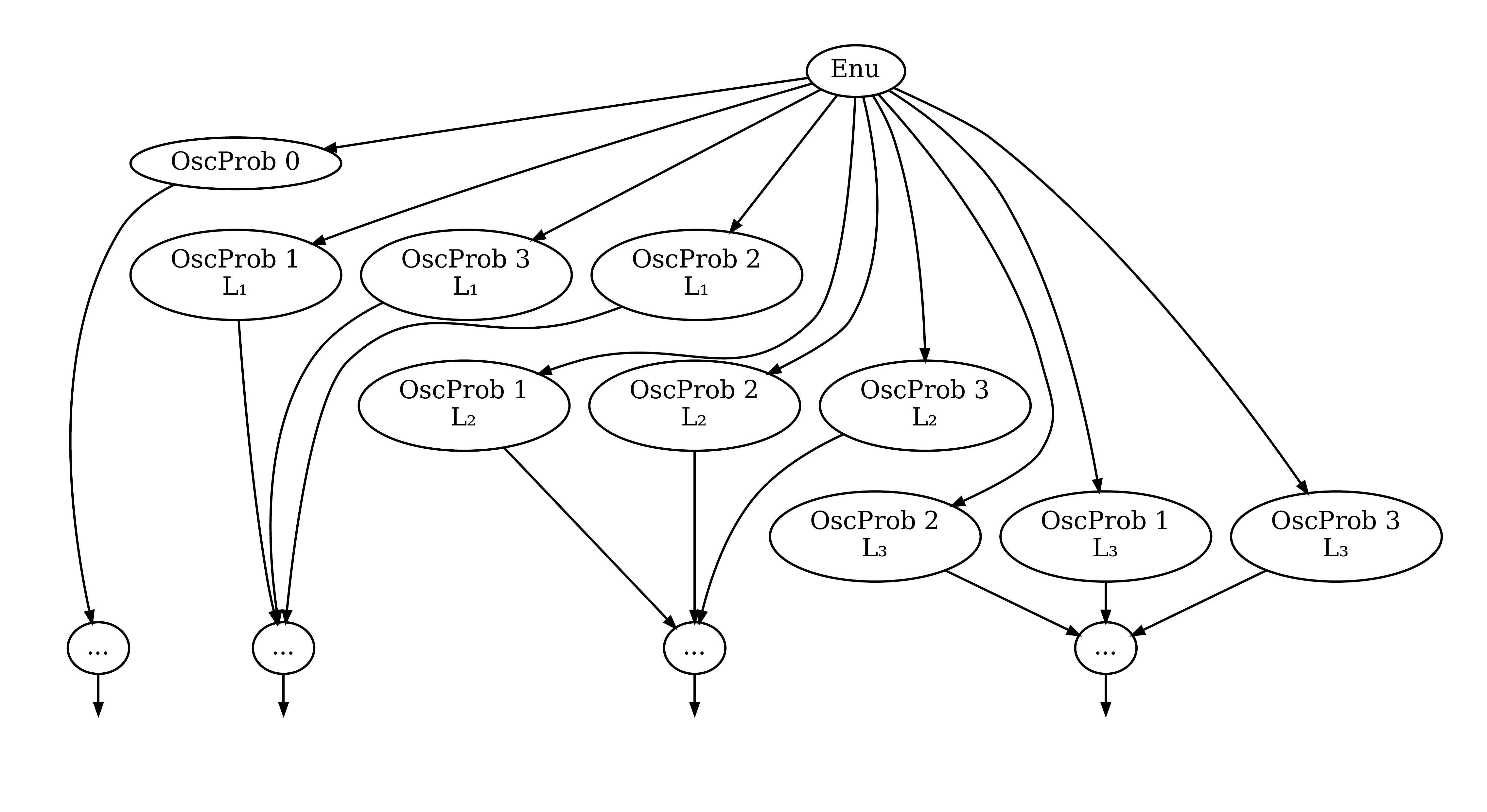}
    \caption{Neutrino oscillation probability calculation scheme. A part of JUNO computational graph.}
    \label{fig:juno}
\end{figure}

Figure~\ref{fig:juno} shows a part of a computational graph for the JUNO experiment implementing the neutrino oscillation
probability calculation (see section~\ref{sec:oscprob}). There are multiple \texttt{OscProb}
transformation instances in the graph computing the neutrino oscillation probability for various
distances $L$,
each of them depending on a vector neutrino energy $\vec{E}_\nu$. For the most practical cases $\vec{E}_\nu$ may be
computed only once. \texttt{OscProb} transformation instances are independent from each other and bound to different parameters
(variables) that may change their output. Parallel technologies are applicable for graphs with such a structure,
since no data writing collision is possible.

The \texttt{OscProb} transformation, as well as most of the framework modules, provide multi-dimensional array operations
which are particularly suitable for multi-threaded systems such as GPUs or multi-core CPUs if their elements are
computed independently.

\section{CUDA overview}
CUDA (Compute Unified Device Architecture) is an architecture for parallel data processing for NVIDIA GPUs. The average
GPU has hundreds of times more threads compared to modern CPUs.
Threads run in parallel in SIMT (Single Instruction, Multiple Threads)~\cite{lindholm2008nvidia} manner as GPUs were
originally created for image processing --- a vivid example of SIMT algorithms.

The CUDA Toolkit~\cite{nvidia2007compute} has a set of specialized libraries optimized for their purposes, such as cuBLAS (linear algebra), cuRAND (random number generators),
cuDNN (deep neural networks), etc. It also provides high-level abstractions to manage computational processes on GPUs, and
low-level methods to tune it.

GPGPU's main performance limitations are memory allocation
and data transfers, as the co-processor is an independent physical device. The copying of data from Host (CPU and
RAM) to Device (GPU) or vice versa is slow. Nevertheless, it
is a powerful tool for accelerating algorithms that contain operations with the same
instruction applied to each element of an array, and producing independent output.

\section{GPU acceleration}
\subsection{Neutrino oscillation probability}
\label{sec:oscprob}
In this section we consider an opportunity of achieving better performance for a distinct transformation that calculates
the neutrino oscillation probability~\cite{giunti2007fundamentals}.

The general formula for oscillation probability in vacuum, the probability that neutrino flavor changes from $\nu_{\alpha}$ to
$\nu_{\beta}$ after travelling distance $L$, reads as follows:
\begin{multline*}
    P(\nu_{\alpha} \rightarrow \nu_{\beta}) = \delta_{\alpha \beta} - 4 \sum_{i>j}\operatorname{Re}(V^*_{\alpha i} V_{\beta i} V_{\alpha j} V^*_{\beta j}) \sin^2 \frac{\Delta m^2_{ij} L}{4E_\nu}+ \\
    + 2 \sum_{i>j}\operatorname{Im}(V^*_{\alpha i} V_{\beta i} V_{\alpha j} V^*_{\beta j}) \sin \frac{\Delta m^2_{ij} L}{2E_\nu},
\end{multline*}
where $E$ denotes neutrino energy, $L$ is a distance between neutrino source and detector, $V_{\alpha i}$ is a complex
unitary matrix called a Pontecorvo-Maki-Nakagawa-Sakata (PMNS) matrix, and $\Delta m^2_{ij} = m_i^2 - m_j^2$ is a neutrino
mass splitting.

Within GNA the oscillation probability is implemented as a set of transformations for each formula item respectively. Each
transformation input is a vector of neutrino energy values $\vec{E}_\nu$.

The computations for different energy values are identical and independent from each other,
therefore they can run in parallel on a GPU. It should be noted that the input array (neutrino energy), in most realistic
cases, is known beforehand and will be copied to the GPU only once while the computation is performed for different
oscillation parameter values.

The following features were used to port the oscillation probability code to GPU:
\begin{itemize}
    \item CUDA Streams~\cite{gomez2012performance},
    \item datasets are divided into smaller sizes to organize overlapped execution,
    \item asynchronous memory copying.
\end{itemize}
After porting the oscillation probability the result was verified: a difference between GPU
and CPU output results is within the roundoff accuracy of the double precision floating point numbers.

\begin{table}[b]
  \centering
  \renewcommand{\arraystretch}{1.5}
  \begin{tabular}{p{7cm}r@{\hspace{1cm}}r}
    \toprule
    Input data size, elements         & $10^4$           & $10^6$           \\
    \midrule
    CPU time / (GPU computing + transfer time)  	& 0.017            & 1.39\phantom{0}  \\
    CPU time / GPU computing-only time 			& 20.90\phantom{0} & 26.46\phantom{0} \\
    \bottomrule
  \end{tabular}
  \vspace{10pt}
  \caption{Benchmarks for oscillation probability calculation on CPU and GPU with input vectors sizes of $10^4$ and
  $10^6$ elements.}
  \label{tab:tab}
\end{table}

Results of the test with input energy vectors of sizes $10^4$ and $10^6$ elements are presented in table \ref{tab:tab}. The
calculation is performed with double precision on Intel Core i7-6700HQ CPU and NVIDIA GeForce GTX 970M GPU.
It should be noted that a size of $10^4$ elements corresponds to the JUNO experiment's case.
First row contains the ratio of the full computation times for CPU-only and GPU-oriented (including data transfer costs)
versions of the algorithm. The second row contains the ratio of the computation times (without data transfer costs in
GPU-based case).

When data transfer is taken into account the acceleration for the $10^6$ sample size is not significant. For the smaller
sample the acceleration is not enough to cover the overhead due to data transfer.

When data transfer is not taken into account the achieved acceleration is at least $\times$20 compared to CPU case.
Since the neutrino energy is computed only once and then stored the latter is the more realistic case for this task.

The speed-up is expected to be more significant for larger datasets. At the same time the data transfer overhead should
be considered and handled appropriately in any case.

It should also be noted that single precision floating point operations are typically much faster (dozens of times)
on most GPUs when compared to double precision. For CPUs the single precision is only twice as faster.
Therefore a significant speed-up is expected for cases when
single precision is sufficient.

\subsection{Computational chains with GPU-oriented transformations}
The original CPU computational scheme was modified in such a way
that switching between CPU- and GPU-oriented transformation modes is transparent for the end user. The transformation is
still a single object with two function definitions: one for the CPU and another for the GPU. On the UI side the GPU
computation is enabled by setting a single flag that changes the target device of the transformation and switches the active
function.
Thus, users are enabled to work with the GPU mode of GNA without any special knowledge about GPGPU.

In order to handle data transfer we implemented a \CC{} wrapper for the GPU array and defined several frequently used
mathematical operations. The portion of the framework that contains CUDA is built as a separate shared library. Then the
main code is built with this library as a dependency. This way GPU functions may be called from the common \CC{} code.
GPU related code may be switched off completely by a special flag during the compilation of the framework.

\begin{figure}[tb]
    \centering
    \tikzset{
  nonterminal/.style={
    rectangle, rounded corners=3pt,
    minimum size=10mm,
    very thick,
    draw=red!50!black!50,         
    top color=white,              
    bottom color=red!50!black!20, 
  },
  terminal/.style={
    rectangle, rounded corners=3pt,
    minimum size=10mm,
    very thick,draw=black!50,
    top color=white,bottom color=black!20,
    font=\ttfamily},
  skip loop/.style={to path={-- ++(0,#1) -| (\tikztotarget)}}
}

{
  \tikzset{terminal/.append style={text height=1.5ex,text depth=.25ex}}
  \tikzset{nonterminal/.append style={text height=1.5ex,text depth=.25ex}}
}

\begin{tikzpicture}[
        point/.style={coordinate},>=stealth',thick,draw=black!50,
        tip/.style={->,shorten >=0.007pt},every join/.style={rounded corners},
        hv path/.style={to path={-| (\tikztotarget)}},
        vh path/.style={to path={|- (\tikztotarget)}},
        text height=1.5ex,text depth=.25ex  
    ]
    \matrix[column sep=4mm, row sep=4mm] {
	& & \node (h2d) [] {H2D}; & & & & & & &  \\
      \node (gpu) [] {GPU:};& & & & \node (t1) [terminal] {$\mathrm{T_1}$}; &  \node (tdots) [] {\ldots}; & \node (tk) [terminal] {$\mathrm{T_k}$};& & & &\\
        \node (cpu) [] {CPU:}; & \node (p1) [point] {}; & \node (ui1) [nonterminal] {$\mathrm{T_0}$};&
        \node (p7) [point] {}; & & & &   &
          \node (ui2) [nonterminal] {$\mathrm{T_{k+1}}$};&
        \node (p9) [point] {};  \node (p10) [point]       {};\\
        & & & & & &  \node (d2h) [] {D2H}; \\
	};
    { [start chain]
        \chainin (p1);
        \chainin (ui1)   [join=by tip];
        \chainin (t1)  [join=by {vh path,tip}];
        \chainin (tdots)    [join=by tip];
        \chainin (tk)    [join=by tip];
        \chainin (ui2)   [join=by {vh path,tip}];
        \chainin (p10)   [join=by tip];
  }
\end{tikzpicture}
    \caption{Schema of mixed (CPU and GPU) computational chain.}
    \label{fig:cpu-gpu-chain}
\end{figure}
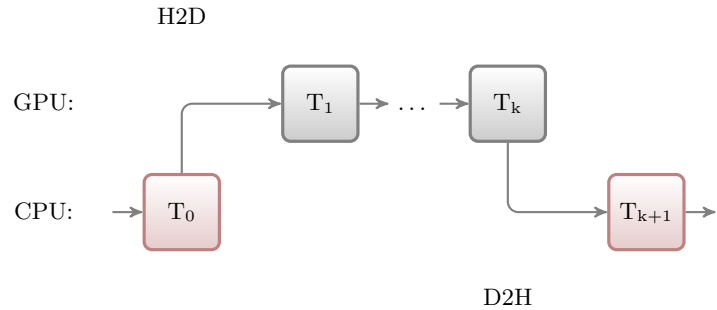

Since memory allocation is one of GPGPU's limitations within GNA, all required memory for both the GPU and CPU is allocated
during the configuration stage to avoid extra time costs in the runtime.

As described earlier, inputs are simply the views on the data of the corresponding outputs of preceding transformations.
The same feature is implemented for the GPU arrays.
There is no additional allocation on the GPU for the inputs as
it refers to the output it is bound to. The only exception to this rule is the first GPU-oriented transformation in the computational
subchain: an extra GPU memory
allocation for its inputs occurs because we need to transfer data from Host memory to
the Device.

We have extended the GNA internal data storage objects in order to maintain a synchronized copy of Host data on the
Device. The synchronization is done in a lazy manner, i.e. it happens only when the unsynchronized Host data is read
from Device and vice versa.

Figure~\ref{fig:cpu-gpu-chain} shows the computation scheme in which the chain contains a subset of GPU-based
transformations. Only two data transfers between the Host and the Device take place in this case: at the beginning of
GPU subchain and at the end of it. We
minimize communication between Host and Device to cut the time costs due to data copying since it is an expensive
operation. The status of GPU function, which indicates whether or not it was executed successfully, is available on the Host side after the
transformation computation is finished. Device-To-Device data transfers may occur inside the transformations
implementation, but they are not considered to be costly.

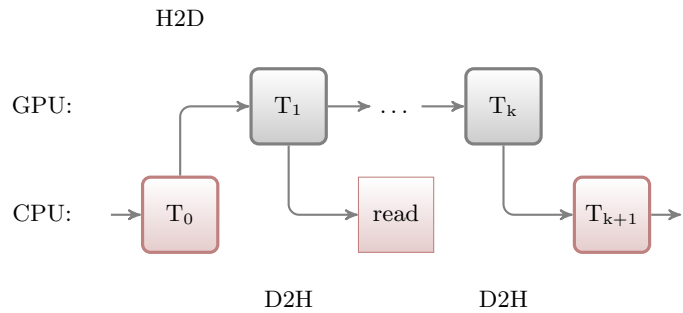
\begin{figure}[tb]
    \centering
    \tikzset{
  nonterminal/.style={
    rectangle, rounded corners=3pt,
    minimum size=10mm,
    very thick,
    draw=red!50!black!50,         
    top color=white,              
    bottom color=red!50!black!20, 
  },
  terminal/.style={
    rectangle, rounded corners=3pt,
    minimum size=10mm,
    very thick,draw=black!50,
    top color=white,bottom color=black!20,
    font=\ttfamily},
  skip loop/.style={to path={-- ++(0,#1) -| (\tikztotarget)}}
}

{
  \tikzset{terminal/.append style={text height=1.5ex,text depth=.25ex}}
  \tikzset{nonterminal/.append style={text height=1.5ex,text depth=.25ex}}
}

\begin{tikzpicture}[
        point/.style={coordinate},>=stealth',thick,draw=black!50,
        tip/.style={->,shorten >=0.007pt},every join/.style={rounded corners},
        hv path/.style={to path={-| (\tikztotarget)}},
        vh path/.style={to path={|- (\tikztotarget)}},
        text height=1.5ex,text depth=.25ex  
    ]
    \matrix[column sep=4mm, row sep=4mm] {
                               &                        & \node (h2d) [] {H2D};                       &                                         &                                                           &                                          &                                                 &                                                 \\
      \node (gpu) [] {GPU:};   &                        &                                             & \node (t1) [terminal] {$\mathrm{T_1}$}; & \node (tdots) [] {\ldots};                                & \node (tk) [terminal] {$\mathrm{T_k}$};  &                                                 &                                                  &  \\
        \node (cpu) [] {CPU:}; & \node (p1) [point] {}; & \node (ui1) [nonterminal] {$\mathrm{T_0}$}; &                                         & \node (read) [nonterminal,thin,rounded corners=0] {read}; &                                          & \node (ui2) [nonterminal] {$\mathrm{T_{k+1}}$}; & \node (p9) [point] {};  \node (p10) [point] {}; \\
                               &                        &                                             & \node (d2h) [] {D2H};                   &                                                           & \node (d2h) [] {D2H};                   \\
	};
    { [start chain]
        \chainin (p1);
        \chainin (ui1)   [join=by tip];
        \chainin (t1)    [join=by {vh path,tip}];
	{ [start branch=read]
		\chainin (read)  [join=by {vh path,tip}];
	}
        \chainin (tdots) [join=by tip];
        \chainin (tk)    [join=by tip];
        \chainin (ui2)   [join=by {vh path,tip}];
        \chainin (p10)   [join=by tip];
  }
\end{tikzpicture}
    \caption{Reading an intermediate result from the GPU chain.}
    \label{fig:gpu-read}
\end{figure}

Extra data transfers from Device to Host may be triggered by the user, reading the data at
any point of the computational chain as is shown in Figure~\ref{fig:gpu-read}. In this case an extra
data transfer occurs. The backward transfer is not needed. Because user-triggered reading may occur during a debugging
procedure or for the plotting of data, the data transfer overhead in not significant in this case when compared to the
actual data analysis.

\section{Future work}

The major shortcoming of the current GPU support implementation is the lack of fault tolerance. In the case of GPU
failure the computation will be aborted.

\begin{figure}[h!]
    \centering
    \tikzset{
  nonterminal/.style={
    rectangle, rounded corners=3pt,
    minimum size=10mm,
    very thick,
    draw=red!50!black!50,         
    top color=white,              
    bottom color=red!50!black!20, 
  },
  terminal/.style={
    rectangle, rounded corners=3pt,
    minimum size=10mm,
    very thick,draw=black!50,
    top color=white,bottom color=black!20,
    font=\ttfamily},
  crossred/.style={
    minimum size=10mm,
    very thick,draw=black!50, cross out,
    top color=white,bottom color=black!20,
    font=\ttfamily},
    cross/.style={
        append after command={
            [every edge/.append style={
                thick,
                red!50!black!50,
                shorten >=\pgflinewidth,
                shorten <=\pgflinewidth,
            }]
           (\tikzlastnode.north west) edge (\tikzlastnode.south east)
           (\tikzlastnode.north east) edge (\tikzlastnode.south west)
        }
    }
}

{
  \tikzset{terminal/.append style={text height=1.5ex,text depth=.25ex}}
  \tikzset{nonterminal/.append style={text height=1.5ex,text depth=.25ex}}
  \tikzset{cross/.append style={text height=1.5ex,text depth=.25ex}}
}

\begin{tikzpicture}[
        point/.style={coordinate},>=stealth',thick,draw=black!50,
        tip/.style={->,shorten >=0.007pt},every join/.style={rounded corners},
        hv path/.style={to path={-| (\tikztotarget)}},
        vh path/.style={to path={|- (\tikztotarget)}},
        text height=1.5ex,text depth=.25ex  
    ]
    \matrix[column sep=4mm, row sep=4mm] {
	& & \node (h2d3) [] {H2D}; & & & & & & &  \\
        \node (gpu3) [] {GPU:};& & &
	\node (t13) [terminal] { $\mathrm{T_1}$}; &
	\node (tdots3) [cross] {\ldots}; &
	\node (tk3) [terminal] {$\mathrm{T_k}$};&
	\node (pgpu3) [point] {}; & &\\
        \node (cpu3) [] {CPU:}; &
	\node (p13) [point] {}; &
	\node (ui13) [nonterminal] {$\mathrm{T_0}$};&
        &
        \node (ui23) [nonterminal] {$\mathrm{T_{1}}$}; &
	\node (tdotscpu3) [] {\ldots}; &
	\node (tkcpu3) [nonterminal] {$\mathrm{T_k}$}; &
 	\node (tkplus13) [nonterminal] {$\mathrm{T_{k+1}}$}; &
        \node (p93) [point] {}; &
	\node (p103) [point] {}; & \\
};
    { [start chain]
        \chainin (p13);
        \chainin (ui13)   [join=by tip];
	{ [start branch=faultgpu]
        	\chainin (t13)  [join=by {vh path,tip}];
        	\chainin (tdots3)  [join=by tip];
        	\chainin (tk3)    [join={by tip, dotted}];
          \chainin (tkplus13)   [join=by {hv path,tip,dotted}];
	}
        \chainin (ui23)   [join=by tip];
        \chainin (tdotscpu3)   [join=by tip];
        \chainin (tkcpu3)   [join=by tip];
        \chainin (tkplus13)   [join=by tip];
        \chainin (p103)   [join=by tip];
  }
\end{tikzpicture}
    \caption{Computational process recovery on CPU after GPU fault.}
    \label{fig:crash}
\end{figure}
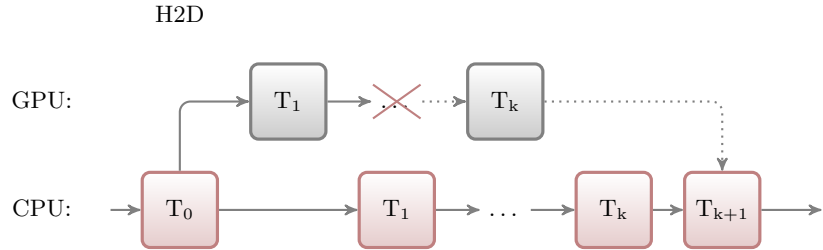

We are planning to add a feature of switching the computation between CPU and GPU modes automatically during runtime as is
shown in Figure~\ref{fig:crash}. It is assumed that the deceleration of the algorithm execution is more preferred than aborting
it.

Another planned feature is adding checkpoints for the GPU side of the framework. It will
decrease latency time for recovering the computation crashed on GPU side. This implies that data will regularly be synchronized between
Host and Device. Since this may lead to an additional overhead the existence and frequency of the checkpoints will be configurable.

In order to use a GPU for the computational chain in a real analysis a subset of existing transformations should be ported to the GPU.
Not every algorithm will be ported, however. The choice will be made based on analysis of the computational chains of the
Daya Bay and JUNO experiments. As a sufficient set of transformations is ported we will benchmark the GPU-enabled
version of GNA on several realistic computational schemes with various configurations and floating point precision
settings.

Since the data transfer costs may negate performance improvement of GPU-enabled computational chain the actual choice of
the configuration should be made and tested by the end-user, based on a particular computational chain. Specialized
benchmarking tools will be implemented in GNA to simplify this task.

\section{Conclusion}

In this paper we describe the GPU support within the GNA framework implemented via the CUDA architecture
with transparency for the end-user. For the particular case of neutrino oscillation probability it has been demonstrated that the
achieved acceleration may be of order of $\times$20 for double precision floating point numbers.

While the realistic acceleration for the large computational chains may be lower and may depend on a particular chain, the
prospects look very promising. Significant improvement is expected when single precision is sufficient for the task.
An acceleration obtained in case of single precision is usually much higher for GPUs compared to CPUs.
The corresponding studies and benchmarks will be performed in further work.

The solutions to the major problems and limitations, such as memory allocation and data transfer are discussed.

\section*{Acknowledgements}
We are grateful to Chris Kullenberg for reading the manuscript and for valuable suggestions.

This research is supported by the Russian Foundation for Basic Research (projects
no.~18-32-00935 and 16-07-00886) and by the Association of Young Scientists and Specialists of Joint Institute for Nuclear
Research (grant no.~18-202-08).

The manuscript has been submitted to ICCSA 2018 (Lecture Notes in Computer Science, publisher: Springer Verlag).

\bibliographystyle{h-physrev.bst}
\bibliography{refs,physics}

\end{document}